\documentclass[aps,pra,10pt,superscriptaddress,twocolumn,superscriptaddress]{revtex4-2}

\usepackage{amsmath, amsthm, amssymb}
\usepackage[]{graphicx} 
\usepackage{dcolumn} 
\usepackage{bm} 
\usepackage[normalem]{ulem}
\usepackage{cases}
\usepackage{color}
\usepackage[dvipsnames]{xcolor}
\usepackage{algorithm}
\usepackage{algpseudocode}
\usepackage{csquotes}
\usepackage{mathtools}
\usepackage{mathrsfs}  
\usepackage{mathtools}
\usepackage{amsmath}
\usepackage[caption=false]{subfig}
\usepackage{lipsum}
\usepackage{comment}
\usepackage{tabularx}
\usepackage{multirow}
\usepackage{braket}
\usepackage{pgf} 
\usepackage{siunitx} 
\usepackage{lipsum}  
\usepackage{booktabs}
\usepackage{hyperref}
\hypersetup{
    colorlinks=true,
    linkcolor=blue,
    citecolor=blue,
    filecolor=magenta,      
    urlcolor=cyan
    }


\usepackage[colorinlistoftodos]{todonotes}

\definecolor{edocolor}{HTML}{008080}
\definecolor{ilektracolor}{HTML}{cc66ff}

\newcommand{\edo}[2]{{\color{edocolor}\sout{#1}#2}}

\definecolor{pinocol}{rgb}{0,.4,1}

\definecolor{mycolorMatteo}{HTML}{00A9CB}

\definecolor{mycolorAntonio}{HTML}{FF1493}

\begin{document}

\title{Intermodal quantum key distribution over an 18 km free-space channel \\ with adaptive optics and room-temperature detectors}

\author{Edoardo~Rossi}
\affiliation{Dipartimento di Ingegneria dell'Informazione, Universit\`a degli Studi di Padova, via Gradenigo 6B, IT-35131 Padova, Italy}

\author{Ilektra~Karakosta-Amarantidou}
\affiliation{Dipartimento di Ingegneria dell'Informazione, Universit\`a degli Studi di Padova, via Gradenigo 6B, IT-35131 Padova, Italy}

\author{Matteo~Padovan}
\affiliation{Dipartimento di Ingegneria dell'Informazione, Universit\`a degli Studi di Padova, via Gradenigo 6B, IT-35131 Padova, Italy}

\author{Marco~Nardi}
\affiliation{Dipartimento di Ingegneria dell'Informazione, Universit\`a degli Studi di Padova, via Gradenigo 6B, IT-35131 Padova, Italy}

\author{Marco~Avesani}
\affiliation{Dipartimento di Ingegneria dell'Informazione, Universit\`a degli Studi di Padova, via Gradenigo 6B, IT-35131 Padova, Italy}

\author{Francesco~B.~L.~Santagiustina}
\affiliation{ThinkQuantum s.r.l., via della Tecnica 85, IT-36030 Sarcedo, Italy}

\author{Marco~Taffarello}
\affiliation{ThinkQuantum s.r.l., via della Tecnica 85, IT-36030 Sarcedo, Italy}

\author{Antonio~Vanzo}
\affiliation{Institute of Photonics and Nanotechnology, National Council of Research of Italy, via Trasea 7, IT-35131 Padova, Italy}

\author{Stefano~Bonora}
\affiliation{Institute of Photonics and Nanotechnology, National Council of Research of Italy, via Trasea 7, IT-35131 Padova, Italy}

\author{Giuseppe~Vallone}
\affiliation{Dipartimento di Ingegneria dell'Informazione, Universit\`a degli Studi di Padova, via Gradenigo 6B, IT-35131 Padova, Italy}
\affiliation{Padua Quantum Technologies Research Center, Universit\`a degli Studi di Padova, via Gradenigo 6A, IT-35131 Padova, Italy}

\author{Paolo~Villoresi}
\affiliation{Dipartimento di Ingegneria dell'Informazione, Universit\`a degli Studi di Padova, via Gradenigo 6B, IT-35131 Padova, Italy}
\affiliation{Padua Quantum Technologies Research Center, Universit\`a degli Studi di Padova, via Gradenigo 6A, IT-35131 Padova, Italy}

\author{Francesco~Vedovato}
\email[Corresponding author: ]{francesco.vedovato@unipd.it}
\affiliation{Dipartimento di Ingegneria dell'Informazione, Universit\`a degli Studi di Padova, via Gradenigo 6B, IT-35131 Padova, Italy}
\affiliation{Padua Quantum Technologies Research Center, Universit\`a degli Studi di Padova, via Gradenigo 6A, IT-35131 Padova, Italy}


\begin{abstract}\noindent
Intermodal quantum key distribution at telecom wavelengths provides a hybrid interface between fiber connections and free-space links, both essential for the realization of scalable and interoperable quantum networks.
Although demonstrated over short-range free-space links, long-distance implementations of intermodal quantum key distribution remain challenging, due to turbulence-induced wavefront aberrations which limit efficient single-mode fiber coupling at the optical receiver.
Here, we demonstrate a real-time intermodal quantum key distribution field trial over an 18 km free-space link, connecting a remote terminal to an urban optical ground station equipped with a 40 cm–class telescope. 
An adaptive optics system, implementing direct wavefront sensing and high-order aberration correction, enables efficient single-mode fiber coupling and allows secure key generation of 200 bit/s using a compact state analyzer equipped with room-temperature detectors.
We further validate through experimental data a turbulence-based model for predicting fiber coupling efficiency, providing practical design guidelines for future intermodal quantum networks.
\end{abstract}

\maketitle
\section{Introduction}
The realization of large-scale quantum networks requires the ability to implement quantum communication protocols across heterogeneous transmission media, like optical fibers and free-space links, which may include satellite channels~\cite{MiciusReview2022, Pittaluga2025, Liu2026, Zheng2026}. 
The interoperability between these media is therefore a central requirement for future network architectures, enabling flexible connectivity, last-mile access, and the integration of terrestrial and space-based terminals~\cite{chen2021integrated, Chen2025_ChineseQuantumNetwork, Li:SatQKD_realtime_2025}. Among the first quantum communication protocols to be deployed outside laboratory environments, quantum key distribution (QKD)~\cite{bennettQuantumCryptographyPublic2014, scaraniSecurityPracticalQuantum2009, sidhuAdvancesSpaceQuantum2021, pirandolaAdvancesQuantumCryptography2020} provides a technologically mature and well-characterized platform for assessing the feasibility of such  intermodal scenarios, where fiber-based and free-space segments are seamlessly integrated to support quantum states propagation through the channel~\cite{piccia_DFA_DEI, Kržič2023, Stathis2026}. 
Furthermore, advancements in quantum communication technology have led to the commercial availability of QKD systems operating at telecom wavelength. However, these systems are generally designed for fiber-based networks, and typically do not tolerate channel efficiency below $-30$~dB~\cite{selentis2026evaluatingrelayedswitchedquantum, alia2024100gbpsquantumsafeipsec, gkouliaras2025demonstrationquantumsecurecommunicationsnuclear}, making them inadequate for the high attenuation encountered in satellite communications.
Therefore, extending these systems to free-space or intermodal links under realistic deployment conditions remains challenging~\cite{avesani2021fulldaylight, gong2018free, liaoLongdistanceFreespaceQuantum2017, Cai:freespace20km2024, bolaños2026ghzratepolarizationbasedqkdfiber}. 

\begin{figure*}
        \centering \includegraphics[width=0.99\textwidth]{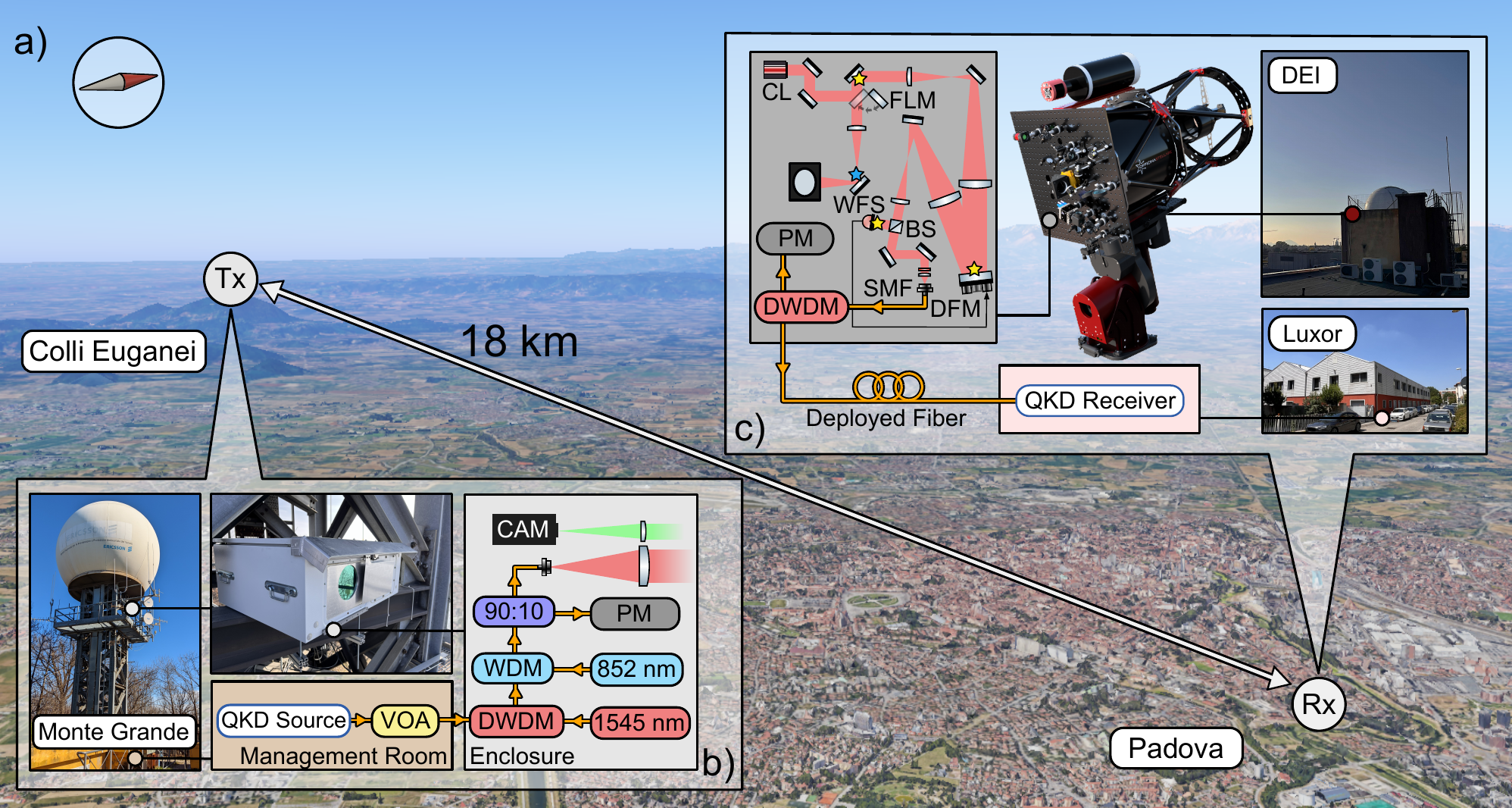}
    \caption{\textbf{Testbed illustration.} \textbf{a} Aerial view of the free-space channel of 18 km. Data from Google Earth [\textcopyright2025 Google]. \textbf{b} Image of the transmitter node located on top of Monte Grande in the Colli Euganei area, including a photo of the enclosure and the sketches of the optical setup. \textbf{c} Image of the receiver node showing the telescope's dome on the DEI rooftop, the 3D illustration of the $0.41$~m telescope equipped with the breadboard, the optical scheme of the AO bench, highlighting the primary focus and pupil planes of reference (blue and yellow stars, respectively), and the photo of the Luxor laboratory where the QKD receiver is placed. VOA: variable optical attenuator; DWDM: dense wavelength-division multiplexer; WDM: wavelength-division multiplexer; 90:10: 90:10 fiber beam-splitter; PM: power meter; CAM: camera; CL: calibration laser; FLM: flip-mirror; DFM: deformable mirror; WFS: wavefront sensor; BS: beam-splitter; SMF: single-mode fiber.}
    \label{fig:setup}
\end{figure*}

In fact, atmospheric turbulence introduces wavefront distortions that severely degrade the coupling efficiency into single-mode optical fibers, which are essential for background suppression, efficient spectral filtering through wavelength-division multiplexing, and compatibility with deployed fiber infrastructure. 
Over low-altitude free-space links longer than a few kilometers, the strength of turbulence often exceeds the regime in which simple angle-of-arrival stabilization is sufficient~\cite{scriminichOptimalDesignPerformance2022, karakosta_balloon}. 
To address these challenges, high-orders adaptive optics (AO) utilize a deformable mirror driven by a sensor's feedback to provide a more complete correction of the turbulence-induced distortions. By operating in a closed-loop configuration, these systems can dynamically compensate the phase aberrations of the received wavefront, though at the cost of a more complex optical setup.

In this work, we demonstrate a real-time intermodal QKD field trial over an $18~\mathrm{km}$ free-space channel connecting a remote optical terminal to an optical ground station (OGS) located in an urban environment. 
We implement an adaptive optics system for single-mode-fiber (SMF) coupling based on direct wavefront sensing and deformable mirror correction, explicitly going beyond mere tip--tilt stabilization, to enable secure key generation even with high overall channel losses. 
By exploiting compact commercial polarization-encoded QKD devices, we  achieve positive secret key rates not only with cryogenic, high-efficiency single-photon detectors, but also by employing detectors operating at room temperature. 
In addition, we expand and validate the turbulence-based model of Ref.~\cite{scriminichOptimalDesignPerformance2022} for predicting SMF coupling efficiency using experimental data, providing design guidelines for interoperable fiber--free-space quantum communication links under realistic atmospheric conditions.

\newpage

\section{Results}

\subsection{Description of the testbed}
\label{sec:setup}

The intermodal QKD testbed combines a long-distance free-space optical link with a deployed fiber connection, as schematically shown in Fig.~\ref{fig:setup}. 
The free-space segment spans 18~km and connects a remote optical transmitter (Tx) located in the Colli Euganei to an OGS situated in Padova. 
The receiving optical terminal (Rx) is equipped with an AO system designed to mitigate atmospheric turbulence and enable efficient coupling of the received quantum signal into a standard SMF.

The quantum channel operates at telecommunication wavelength and relies on commercially available polarization-encoded QKD devices (QUKY platform by ThinkQuantum) originally developed for fiber-based operation. 
The quantum signal is emitted at a wavelength of $\lambda_{\rm QKD} = 1565.50$~nm (ITU channel~15). 
The transmitted beams propagate over the 18~km horizontal free-space channel and are collected at the receiving OGS by a telescope (PRORC400 by Officina Stellare) with an aperture diameter of $D_{\rm Rx} = 410$~mm. 
To support long-distance free-space  propagation and adaptive optics operation at the receiver, the quantum signal is multiplexed at the transmitter's side with auxiliary optical beacons, as described in more details in Section~\ref{sec:TX_node}. 

The OGS is located at the Department of Information Engineering (DEI) of the University of Padova.
After being collected by the telescope, the incoming beams are directed to the AO bench, which employs direct wavefront sensing via a Shack-Hartmann wavefront sensor (WFS) and deformable mirror (DFM) correction, explicitly extending beyond simple angle-of-arrival (tip--tilt) compensation in order to correct higher-order aberrations that significantly impact coupling efficiency. Both the WFS and the DFM are provided by Dynamic Optics. 

\begin{figure*}
    \centering\includegraphics[width=0.99\textwidth]{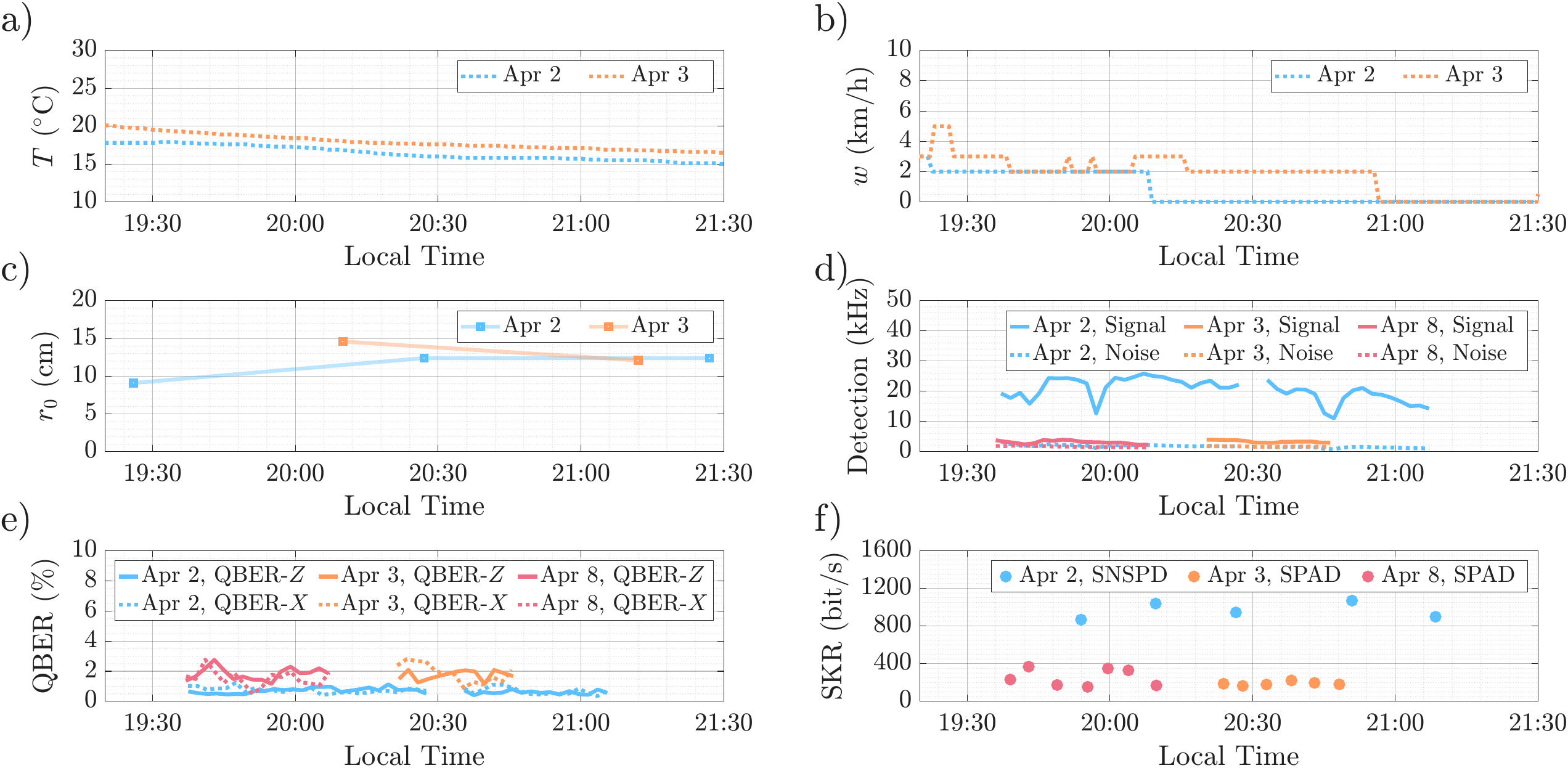}
    \caption{\textbf{QKD results and atmospheric conditions during experiments.} \textbf{a} Temperature, \textbf{b} wind velocity and \textbf{c} Fried parameter measured on two consecutive days during QKD experiments with SPADs and SNSPDs. \textbf{d} Signal and noise dection rate, \textbf{e} QBER and \textbf{f} secret key rate generated during the experiments. }
    \label{fig:QKDresults}
\end{figure*}

Following adaptive-optics-assisted correction, the quantum signal is injected at the OGS into a standard SMF, providing efficient spatial filtering and compatibility with deployed fiber infrastructure. 
After fiber coupling, a dense wavelength-division multiplexer (DWDM) with 100~GHz spacing is used to separate the quantum channel from the beacon signal. 
The quantum signal is then routed through a deployed fiber link of approximately 0.5~km to the QKD receiver located at Luxor laboratories. 
We stress that the OGS does not need to operate as a trusted node, as the QKD receiver is located in a physically separate building with respect to the telescope.

The QKD receiver platform allows operation with different detector technologies, including external high-efficiency superconducting nanowire single-photon detectors (SNSPDs) and intenral room-temperature InGaAs single-photon avalanche diodes (SPADs). 
The overall channel attenuation $\eta_{\rm CH}$ is defined from the Tx output to the input of the QKD receiver and includes free-space propagation losses, optical losses within the receiving terminal, single-mode fiber coupling efficiency, and fiber transmission losses, as specified in Sections~\ref{sec:Channel_Model} and~\ref{sec:AOdesign}. We refer to Sections~\ref{sec:TX_node} and~\ref{sec:Rx_node} for additional implementation details.

In the following, we present the QKD results demonstrating secure key generation at overall channel losses of approximately 30 dB.
Moreover, from the data collected with the WFS, we will show that the model of Section~\ref{sec:AOdesign}, which is derived from Ref.~\cite{scriminichOptimalDesignPerformance2022}, provides reliable estimates of the Fried parameter $r_0$~\cite{fried1965statistics}, which is a measure of the turbulence strength of the link, and of the corresponding expected value for the coupling efficiency into the SMF.

\subsection{Quantum Key Distribution results}
\label{sec:QKD_results}

The QKD platform used in the experiment exploits polarization encoding to implement the 3-state 1-decoy efficient BB84 protocol~\cite{Rusca2018}.
Synchronization between Alice and Bob is achieved through a qubit-based stra\-tegy for clock and offset recovery called Qubit4Sync~\cite{Calderaro2020}. 
The classical channel used for real-time post-processing is implemented by establishing a virtual local area network (LAN) between the QKD source and the QKD receiver, leveraging Internet connectivity on both the Monte Grande and Luxor sides through a dedicated virtual private network (VPN). The QKD experiment was conducted on three separate occasions during April 2025. 

In the first run on April 2\textsuperscript{nd}, we employed external SNSPDs by IDQuantique with 80\% efficiency, while for the second one on April 3\textsuperscript{rd} we used the internal SPADs of the QKD Receiver, characterized by a detection efficiency of 15\%. 
A further run with SPADs was carried out on April 8\textsuperscript{th} to confirm the repeatability of the measurements and the robustness of the experimental platform. 
In addition, during the QKD experiments on April 2\textsuperscript{nd} and 3\textsuperscript{rd}, we monitored key atmospheric parameters relevant to characterizing the free-space link --- namely the Fried parameter $r_0$, temperature $T$, and wind speed $w$ --- as shown in Figs.~\ref{fig:QKDresults}a--\ref{fig:QKDresults}c. 
The temperature and wind speed values were extracted from the DEI weather station logs, while the estimation of $r_0$ was derived from WFS data acquired with the AO system turned off (an example is shown in Fig.~\ref{fig:AOresults}a for the measurement at 19:26 on April 2). 
We note that during the almost hour-long experiment with SNSPDs, we temporarily paused the QKD transmission around 20:30 to record an additional $r_0$ measurement in the middle of the experiment. 

For the SPAD-based trial on April 3\textsuperscript{rd}, we estimated $r_0$ at both the beginning and end of the session. From the data shown in Figs.~\ref{fig:QKDresults}a--\ref{fig:QKDresults}c, we observe that the atmospheric conditions on April 2\textsuperscript{nd} and 3\textsuperscript{rd} were similar. 
As a result, we can deduce that the performance of the AO system, which depends on the turbulence strength, as will be explained in Section~\ref{sec:AOdesign}, was analogous during both runs, thus allowing a fair comparison of system performance using different types of detectors.

Fig.~\ref{fig:QKDresults}d shows the signal and noise detection rates obtained during the different QKD runs. 
These rates are extracted from the logs saved in real-time by the QKD platform. 
To filter out the noise from the signal, a temporal window of $\Delta t = 600$~ps is set by the QKD software around the expected peak in the histogram of arrival time of the incoming pulses. 
Such a histogram is reconstructed by applying the Qubit4Sync algoritm in real-time along the QKD run. 
The average  of the signal rate in the case of SNSPDs and SPADs is around 20.4~kHz and 3.4~kHz, respectively. 
Both these signal rates correspond to a channel efficiency close to $\eta_{\rm Ch} = -29$~dB when rescaled for the different detection efficiencies and the internal optical losses of the QKD receiver of $1.2$~dB. 

Regarding the average noise rate it is possible to see from Fig.~\ref{fig:QKDresults}d that it remains around 2~kHz for both implementations. 
Fig.~\ref{fig:QKDresults}e shows the measured quantum-bit-error-rate (QBER) in the two measurement bases $Z$ and $X$. 
Notably, the average QBER with SNSPDs is below 1\%, while with SPADs is around 2\%. 
The secret key rate (SKR) generated during the different runs, shown in Fig.~\ref{fig:QKDresults}f, is obtained in real-time by the post-processing software installed on the QKD devices.
Such software follows the specification for the used protocol and incorporates finite-size effects on sifted key blocks of size $n_Z$ according to Ref.~\cite{Rusca2018}, in which we set $n_Z$ to be 250000 (50000) bytes in the case of SNSPDs (SPADs). 
As expected, the SKR obtained with SNSPDs (around 1~kbit/s on average) is greater than the one obtained with SPADs (around 200~bit/s).

It is worth noting that the plug-and-play QKD platform employed in this field test --- while originally designed and validated for fiber-based communication --- operates with the intermodal free-space link without requiring functional modifications. 
This result enables broader implementations of QKD networks, as it demonstrates that the presence of the free-space segment does not limit the system’s performance if adaptive optics is exploited.

\subsection{Adaptive optics system data analysis and estimate of fiber-coupling efficiency}
\label{sec:AO_results}

In this Section, we focus on quantifying the performance of the AO system with respect to the turbulence parameters characterizing the optical link, estimated from the WFS data acquired by the AO control software PhotonLoop~\cite{JMocci_Photonloop}. In particular, assuming a Kolmogorov power spectral density for the refractive index fluctuations, it is possible to express the variance associated to each aberration term due to turbulence as~\cite{noll1976zernike, boremanZernikeExpansionsNonKolmogorov1996}
\begin{equation}
    \sigma_{j, {\rm turb}}^{2} = \left( \frac{D_{\rm Rx}}{r_0} \right) ^{\frac{5}{3}} g(j) 
    \label{eq:eq_fit}
\end{equation}
where, as further detailed in Section~\ref{sec:AOdesign}, $g(j)$ is a function that depends on the radial order of the specific aberration expressed through the standard Zernike polynomial.

\begin{figure*}
    \centering \includegraphics[width=0.85\textwidth]{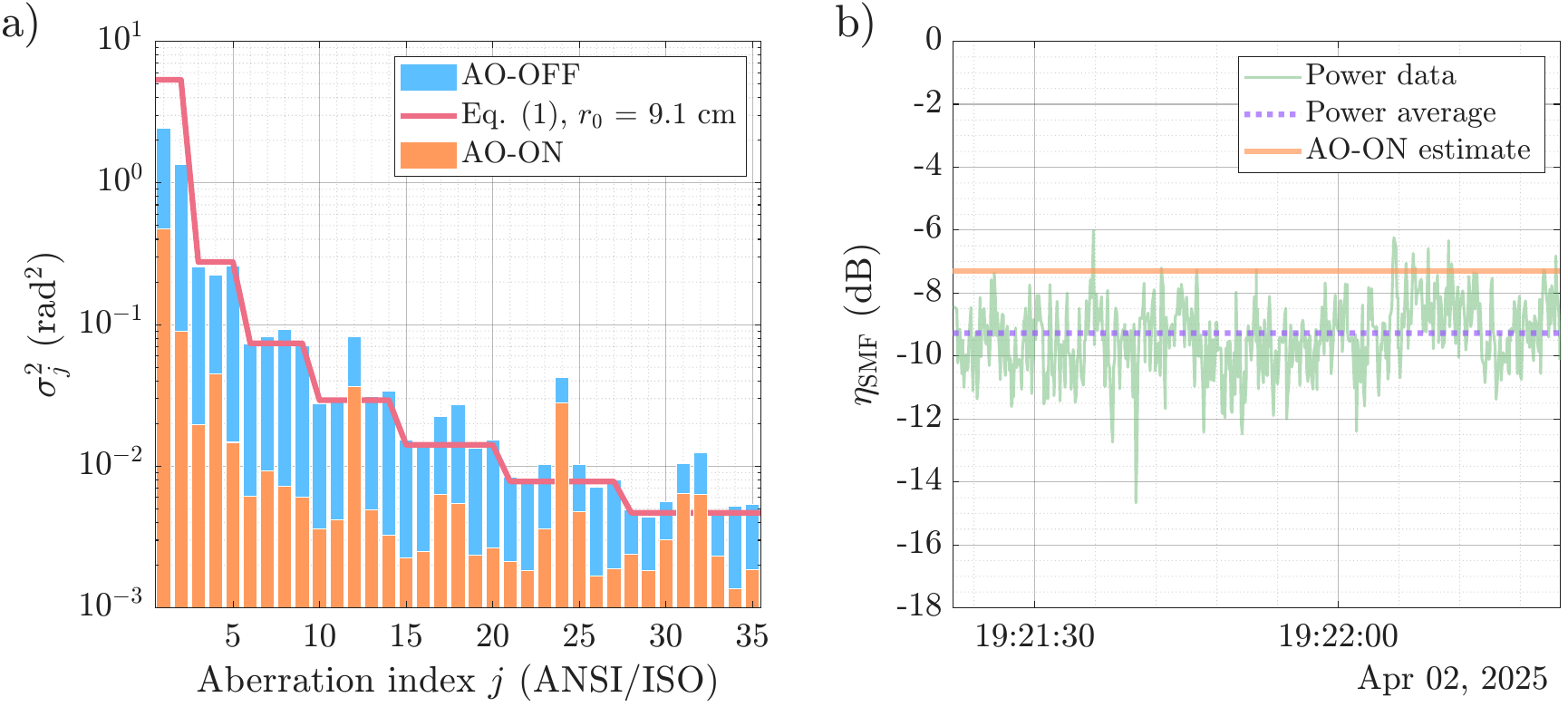}
    \caption{\textbf{Results of the adaptive optics system.} \textbf{a} Aberration variances of Zernike modes with AO-OFF/ON. In the absence of AO correction, the aberrations induced by turbulence follow the Kolmogorv spectrum as described by Eq.~\eqref{eq:eq_fit}.  \textbf{b} Comparison between the coupling efficiency estimated from wavefront sensor data and the corresponding values obtained from direct power measurements.}
    \label{fig:AOresults}
\end{figure*}

Without introducing any correction with the AO system (AO-OFF), it is possible to experimentally estimate the Fried parameter that characterizes the turbulence of the channel from the time-series of the Zernike coefficients measured by using the WFS. 
As an example, we report in Fig.~\ref{fig:AOresults}a a data acquisition taken on April 2\textsuperscript{nd} 2025, around 19:20 CEST. Remarkably, the data obtained with AO-OFF are well aligned with the expected phase aberrations due to turbulence given by Eq.~\eqref{eq:eq_fit} (solid red line), resulting in a more confident estimate of the Fried parameter $r_0$. 

From the recorded variances and the estimation of the Fried parameter it is also possible to infer the expected value of the coupling efficiency when the control is on (AO-ON). 
In particular, for an optical system with at least a first-order (tip-tilt) AO correction stage, the average coupling efficiency into an SMF can be modeled as a product of three multiplicative terms~\cite{scriminichOptimalDesignPerformance2022} 
\begin{equation}
    \eta_{\rm SMF} = \eta_{0} \eta_{S} \eta_{\rm AO} \ ,
    \label{eq:eta_SMF}
\end{equation}
where $\eta_0$ is the optical efficiency of the receiver, $\eta_S$ is the contribution due to scintillation, and  $\eta_{\rm AO}$ is the efficiency of the AO system in correcting the turbulence-induced phase perturbations of the wavefront. 

The AO efficiency term, $\eta_{\rm AO}$, can be further decomposed into two multiplicative components: 
\begin{equation}
    \eta_{\rm AO} = \eta_{\varphi} \eta_{\tau} \ ,
    \label{eq:eta_AO}
\end{equation}
where $\eta_{\varphi}$ quantifies the performance of the AO system in reducing the spatial variance of the phase perturbations and $\eta_{\tau}$ characterizes its temporal response to dynamic wavefront aberrations~\cite{tyson2022principles}. 
Moreover, to estimate the spatial efficiency $\eta_{\varphi}$, it is possible to decompose the phase variance into two contributions: 
\begin{equation}
    \eta_{\varphi} = \eta_{\varphi, {\rm ON}} \eta_{\varphi(J)} \ ,
    \label{eq:eta_varphi}
\end{equation}
where $\eta_{\varphi, {\rm ON}}$ represents the aberrations effectively corrected by the AO system, while $\eta_{\varphi(J)}$ corresponds to uncorrected higher-order modes, with $J$ being the maximum Zernike term corrected by the AO system. 
Therefore, by exploiting the experimental variances $\sigma_j^2$ derived from the WFS data with control on, it is possible to compute the average AO spatial efficiency of the first $j<J$ modes by~\cite{scriminichOptimalDesignPerformance2022}
\begin{equation}
    \eta_{\varphi, {\rm ON}} = \prod_{j = 1}^{J} \frac{1}{\sqrt{1+2 \sigma_{j}^2}} \ ,
\end{equation}
where we set $J=35$. The contributions coming from the uncorrected higher order terms ($j>J$) follow the behavior illustrated in~\cite{noll1976zernike}, and are proportional to $D_{\rm Rx}/r_0$, as detailed in Section~\ref{sec:AOdesign}.

Regarding the second term in Eq.~\eqref{eq:eta_AO}, $\eta_\tau$, the impact of the temporal characteristics of the turbulence and the AO system on the coupling efficiency depends on the Greenwood frequency, which represents the rate at which phase distortions in the wavefront evolve due to wind motion~\cite{Greenwood:77}. For our horizontal link the rejection bandwidth of the AO system~\cite{tyson2022principles} is limited to 10 Hz due to USB communication between the WFS and PhotonLoop. 
Hence, in the example presented in Fig.~\ref{fig:AOresults}a, from the data provided by the WFS we obtain $ \eta_{\varphi, {\rm ON}} = -2.8$~dB and $\eta_{\varphi(J)} = -0.7$~dB, yielding a value for Eq.~\eqref{eq:eta_varphi} of $\eta_\varphi = -3.5$~dB. 
The wind speed measured from the weather station of DEI ($w=2$ km/h at 19:20, see also Fig.~\ref{fig:QKDresults}a) combined with the estimate obtained for the Fried parameter, yields a temporal AO correction efficiency of $\eta_\tau = -0.5$~dB, for a total efficiency of the AO system of $\eta_{\rm AO} = -4$~dB. 
From the remaining terms in Eq.~\eqref{eq:eta_SMF}, further described in Section~\ref{sec:AOdesign}, we obtain that the optical efficiency is $\eta_0 = -2.7$ dB, and the contribution from scintillation is $\eta_S = -0.5$~dB. 
Thus, by putting all the terms together, we estimate a SMF coupling efficiency of $\eta_{\rm SMF} = -7.2$~dB, as shown by the solid orange line in Fig.~\ref{fig:AOresults}b, which is equivalent to 19\%.

At the same time, we can directly obtain the coupling efficiency by measuring the ratio of the power into the fiber $P_{\rm In}$ over the power in front of it $P_{\rm Front}$. 
The power into the fiber is measured by acquiring the C40 beacon via the fiber-coupled power meter (PM) connected to the reflection port of the DWDM at the receiver, while the power in front of the fiber can be derived from the power $P_{\rm Focus}$, measured with a free-space PM at the receiving primary focal plane and considering the fixed losses from the focus to the fiber. 
For the reported example on April 2\textsuperscript{nd} 2025, we acquired data of $P_{\rm In}$ and $P_{\rm Focus}$ at 19:20 CEST, just before running the QKD experiment, to directly measure the coupling efficiency (see solid green line Fig.~\ref{fig:AOresults}b). 
The average value of this direct measurement of coupling results in an efficiency of $\eta_{\rm SMF}=-9.2$~dB (12\%), thus being within 2~dB with respect to the value estimated from the WFS data. 
The discrepancy between the measured fiber coupling and the estimated value can be attributed to non perfect alignment caused by manual coupling optimization into the fiber. Moreover, when the lenslet intensity falls below a predefined threshold, PhotonLoop flags the corresponding centroid estimate as invalid and excludes it from the centroids vector, thereby removing those entries from subsequent matrix computations and leading to a loss of information for wavefront reconstruction~\cite{JMocci_Photonloop}.

\section{Discussion}
We have demonstrated real-time intermodal free-space QKD at telecommunication wavelength over an 18~km ground-to-ground low-altitude link, effectively exploiting adaptive optics in a 40~cm-class telescope to mitigate atmospheric turbulence.
In particular, we achieved a mean secret key rate of 1~kbit/s with high-efficiency SNSPDs and, notably, 200~bit/s with room-temperature SPAD detectors.
These results demonstrate the practical viability of long-range urban free-space quantum communication using sub-meter-class telescopes together with commercially available plug-and-play QKD devices.
Moreover, exploiting the direct measurements of the Fried parameter along the link with WFS data, we obtained a robust estimate of the achievable single-mode-fiber coupling efficiency which is aligned with the average experimental value.
This agreement provides an additional validation of our free-space system design methodology~\cite{scriminichOptimalDesignPerformance2022}, previously established only for short-range horizontal links~\cite{vedovatoRealizationIntermodalFiber2023, piccia_DFA_DEI, bolaños2026ghzratepolarizationbasedqkdfiber}.

By enabling a single OGS to serve multiple users through intermodal links, operating as a transparent optical relay without performing quantum measurements, this QKD architecture at telecom wavelength reduces the need for redundant devices required for trust-node implementations~\cite{chen2021integrated, Chen2025_ChineseQuantumNetwork, picciariello2024quantum}. Moreover, although QKD is used here as the reference quantum communication protocol, the demonstrated intermodal architecture and adaptive-optics-assisted single-mode fiber coupling are protocol-agnostic and thus directly applicable to broader quantum networking tasks, such as entanglement distribution for delegated computing, sensing and communication \cite{delegated_broadbent,sensing,de2023satellite}.

While the implemented adaptive optics system significantly enhances single-mode-fiber coupling under weak-to-moderate turbulence conditions, its performance remains limited in strong-turbulence regimes, such as those typically encountered in long-distance daytime operation.
Addressing this limitation will require adaptive optics architectures with higher rejection bandwidths, capable of compensating more rapidly  the evolution of wavefront distortions.
In particular, replacing software- and interface-limited control schemes with dedicated real-time controllers for wavefront sensing and deformable mirror actuation is expected to provide substantially improved performance under strong turbulence, enabling long-range daylight QKD.

Beyond the specific field trial reported here, the combination of commercial QKD devices with wavefront-sensing-based adaptive optics has broader implications for next-generation quantum networks. By enabling efficient and interoperable fiber--free-space operation under realistic deployment conditions, intermodal links provide a practical building block for satellite-based quantum communication infrastructure.
In this context, the demonstrated performance and system architecture are directly relevant to forthcoming European Space Agency missions such as Eagle-1 and SAGA, which aim to validate QKD services via satellite-to-ground links and to establish a pan-European quantum communication capability~\cite{SES_Quantum_EAGLE1_2024, ESA_SAGA_2025, Hiemstra_2025}.

\section{Methods}

\subsection{Channel Model}
\label{sec:Channel_Model}
In our intermodal link, we express the total channel efficiency $\eta_{\rm Ch}$ ---  from the optical transmitter's aperture at Monte Grande to the input fiber interface of the QKD receiver at Luxor --- as the product
\begin{equation}
    \eta_{\rm Ch} = \eta_{\rm Focus} \eta_{\rm Optics} \eta_{\rm SMF} \eta_{\rm Fiber} \ , \label{eq:etaCh}
\end{equation}
where $\eta_{\rm Focus}$ is the transmission efficiency from the transmitter's aperture to the primary focus of the receiving telescope, $\eta_{\rm Optics} = -4.5$~dB corresponds to the optical efficiency due to imperfect reflectivity and transmittivity of the AO bench from the primary focus to the SMF entrance, $\eta_{\rm Fiber} = -2.4$~dB refers to the transmission efficiency of the fiber between DEI and Luxor, and  $\eta_{\rm SMF}$ is the coupling efficiency already described in Section~\ref{sec:AO_results}. 
We typically measure the link efficiency up to the primary focus $\eta_{\rm Focus}$ in the first steps of the alignment procedure, obtaining a value that ranges from $-17$ to $-10$~dB. 
This value is in agreement with the following model for beam propagation along a horizontal channel.

The channel efficiency up to the primary focus $\eta_{\rm Focus}$ of the receiving optical system can be modeled as the product of the link efficiency due to atmospheric absorption $\eta_A$, and by the collection efficiency of the receiving telescope $\eta_{\rm Coll}$, i.e.,
\begin{equation}
    \eta_{\rm Focus} = \eta_A \eta_{\rm Coll} \,.
\end{equation}

In principle, these terms can be evaluated singularly. For example, the absorption efficiency 
\begin{equation}
   \eta_A = e^{-A(\lambda) L} 
\end{equation}
depends on the absorption coefficient $A(\lambda)$ (which is wavelength-dependent and whose magnitude is of the order of $0.1-0.3$~dB/km for the design wavelength~\cite{scriminichOptimalDesignPerformance2022}) and on the link distance $L$, thus yielding an absorption efficiency $\eta_A$ in the range $-6$ to $-1$~dB for our 18 km-long link. 

\begin{figure}
    \centering    
    \includegraphics[width=0.9\linewidth]{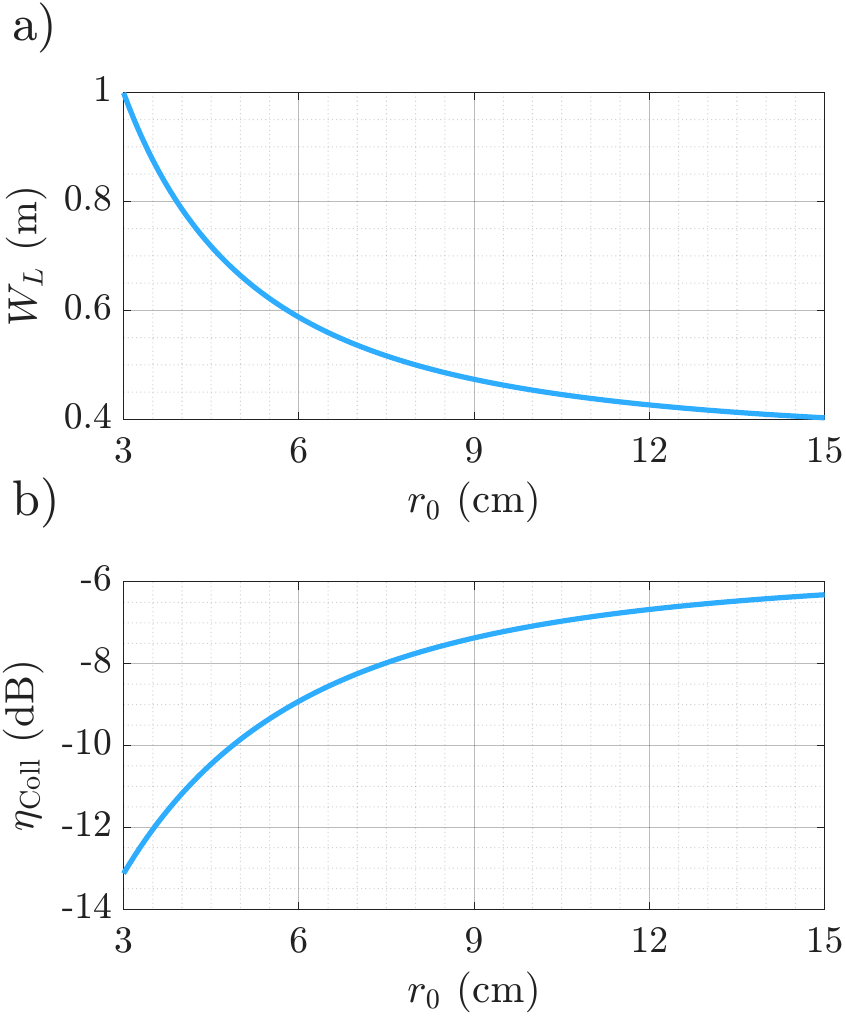}
    \caption{\textbf{Channel parameters estimation due to turbulence:} \textbf{a} Expected waist for the received beam after 18 km of free-space propagation. \textbf{b} Expected collection efficiency for our telescope of $D_{\rm Rx}$ aperture with an obstruction of $D_{\rm Obs}$.}
    \label{fig:etacoll}
\end{figure}

The collection efficiency can be estimated via \cite{Bonato2009}
\begin{equation}
    \eta_{\rm Coll} =  \eta_{\rm Tel} \left[\exp\left(-\frac{D_{\rm Obs}^2}{2W_L^2}\right) - \exp\left(-\frac{D_{\rm Rx}^2}{2W_L^2}\right)\right]
    \label{eq:eta_coll}
\end{equation} 
where $\eta_{\rm Tel} = -1.4$~dB is the reflection efficiency of the telescope, $W_L$ represents the beam radius at the aperture (i.e., located at a distance $L$ from the waist), and $D_{\rm Obs} = 168$~mm is the diameter of the central obstruction. The received beam waist $W_L$ can be related to the Fried parameter $r_0$ via $ W_L = \theta L$, where the half-angle divergence 
\begin{equation}
 \theta = \sqrt{\theta_0^2 + \theta_{\mathrm{turb}}^2}
\end{equation}
includes the intrinsic divergence $\theta_0 = \lambda/(\pi W_0)$  and the beam broadening due to turbulence $\theta_{\rm turb} = \lambda / (\pi \rho_0)$, with $\rho_0 = r_0/2.1$ the spatial coherence radius~\cite{Andrews_book}. According to the simulation reported in Fig.~\ref{fig:etacoll}, we expect a collection efficiency $\eta_{\rm Coll}$ ranging from -13 to -6~dB. Remarkably, the model provides a received beam waist $W_L$ ranging from 0.4~m to 1~m, which is well aligned with our experimental observations.

\begin{figure*}
    \centering \includegraphics[width=0.95\textwidth]{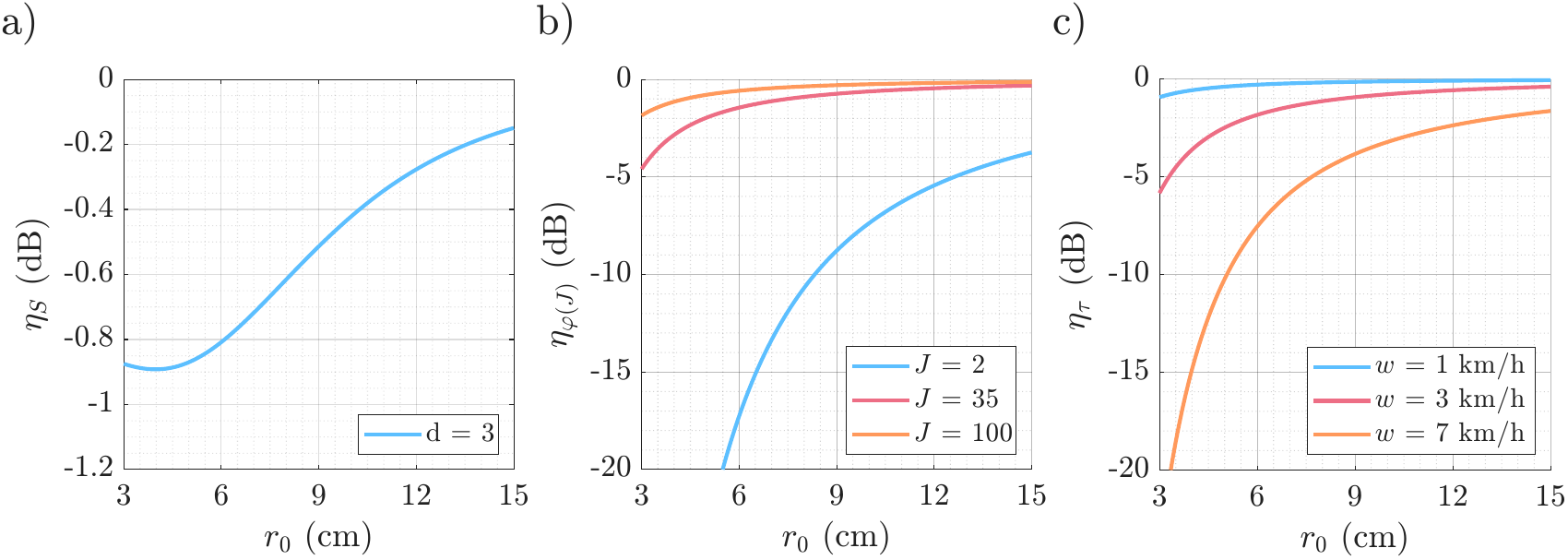}
    \caption{\textbf{Simulations of the different efficiency terms contributing to $\eta_{\rm SMF}$}. Common parameters used in the simulations are the link distance $L=18$~km, the receiver telescope diameter $D_{\rm Rx} = 410$~mm, and the wavelength $\lambda = 1555$~nm. We vary the Fried parameter $r_0$ in the range 3 to 15~cm, as expected in the real experiment, corresponding to a $D_{\rm Rx}/r_0$ ratio between 3 and 14. \textbf{a} Efficiency $\eta_S$ due to scintillation as a function of the Fried parameter $r_0$. \textbf{b} Efficiency $\eta_{\varphi(J)}$ due to residual spatial phase variance after having perfectly corrected for $J$ modes ---  Eq.~\eqref{eta:spatial:J} --- as a function of the Fried parameter $r_0$. Case $J=2$ implies just tip-tilt correction, $J=35$ is our design baseline, and $J=100$ is shown for comparison. \textbf{c} Efficiency $\eta_\tau$ as a function of the Fried parameter $r_0$ by assuming different wind conditions.}
    \label{fig:simulations}
\end{figure*} 

\subsection{Design considerations for the Adaptive Optics system}
\label{sec:AOdesign}

In the following, we describe the design of the AO system used in the experiment, with particular emphasis on the factors that, in addition to AO performance, affect the SMF coupling efficiency. 
We adopted a design wavelength of $\lambda = 1555$~nm, as it lies between the operational wavelengths $\lambda_{\rm Beacon}$ and $\lambda_{\rm QKD}$ employed in the experiment. 

As discussed in Section~\ref{sec:AO_results}, the average single-mode coupling efficiency, given by Eq.~\eqref{eq:eta_SMF}, can be factorized into three contributions. 
The mode-mismatch term $\eta_0$, expressed as
\begin{equation}
    \eta_0 = 2\left[\frac{\exp(-\beta^2) -\exp(-\beta^2\alpha^2)}{\beta\sqrt{1-\alpha^2}} \right]^2\ ,
\end{equation}
mainly depends on the linear obscuration ratio $\alpha = D_{\rm Obs}/D_{\rm Rx} = 0.41$ and on the effective focal length $f_{\rm Eff} \approx 2$~m of the receiving system, with
\begin{equation}
    \beta = \frac{\pi D_{\rm Rx} \rm}{4\lambda} \frac{\rm MFD}{f_{\mathrm{Eff}}} 
    \label{eq:beta} \ .
\end{equation}
It is important to note that $\eta_0$ represents a mode-overlap integral and is therefore independent of atmospheric turbulence~\cite{Wang2018_coupling_eta0}. 
Therefore, even in absence of turbulence, the optical efficiency for coupling the received signal into a standard SMF with a mode-field diameter (MFD) of 10.4~$\mu$m is fundamentally limited by the telescope design. 
For this system, the maximum achievable optical efficiency $\eta_0$ results in $-2.6$~dB.
Achieving this value would require an optimal effective focal length, which could be realized through customized optical elements specifically designed for the system. 
However, aiming to design a system based, as much as possible, on commercially available components, we selected standard lenses with non-optimal focal lengths. 
This choice results in an effective focal length $f_{\rm Eff}$ corresponding to a mode-matching factor $\beta = 1.1$. 
Consequently, the optical efficiency of the system is $\eta_0 = -2.7$~dB, which remains very close to the maximum achievable.

The SMF coupling efficiency due to scintillation $\eta_S$ can be evaluated by approximating the received wave as a spherical wave. 
This last assumption is well placed since the Rayleigh range $z_0 = \pi W_0^2/\lambda$ of the transmitted beams is around 1.3~km, which is way smaller than the link distance. 
The efficiency due to scintillation can be evaluated via~\cite{scriminichOptimalDesignPerformance2022}
\begin{equation}
    \eta_S = e^{-\sigma_\chi^2} \label{eq:etaS} \ ,
\end{equation}
where 
\begin{equation}
    \sigma_\chi^2 = \frac{1}{4}\mathrm{ln}\left(\sigma_I^2 + 1\right)
\end{equation}
is the log-amplitude variance for a spherical wave, while $\sigma_I^2$ is the scintillation index~\cite{Andrews_book}. 
In our case, the size of the receiving aperture is larger than the correlation width $\rho_c$, which is given by~\cite{fante}
\begin{align}
    \begin{split}
        \rho_c^{(\text {weak})} &=\sqrt{\lambda L} \\ 
        \rho_c^{\text {(strong)}} &=0.36 \sigma_R^{-3 / 5} \sqrt{\lambda L}
    \end{split}    
\end{align}
for weak and strong turbulence, where $\sigma_R^2$ is the Rytov variance
\begin{equation}
    \sigma_{\mathrm{R}}^2=1.23 C_{\mathrm{n}}^2 k^{7 / 6} L^{11 / 6}
\end{equation}
with $C_n^2$ the refractive index structure constant, $L$ the link distance and $k=2\pi/\lambda$ defined as the wave number ~\cite{Andrews_book}. 
It is worth to note that, by knowing the value of the Fried parameter $r_0$, it is possible to derive the value of the refractive index structure constant $C_n^2$, since for a spherical wave propagating over a horizontal link one has~\cite{Andrews_book, tyson2022principles}
\begin{equation}
r_0 = \left( 0.16 \ C_n^2 k^2 L \right)^{-3/5} \ . \label{eq:r0}
\end{equation}
Since the aperture at the receiver is larger than the correlation width, there is a reduction in scintillation due to aperture averaging, which is expressed through the aperture-averaged scintillation index
\begin{equation}
    \sigma_I^2\left(D_{\rm Rx}\right) = \exp\left( T_1 + T_2 \right)-1 \ ,
\end{equation}
with
\begin{align}
    T_1 &= \frac{0.49 \beta_0^2}{\left(1 + 0.18 d^2 + 0.56 \beta_0^{12/5}\right)^{7/6}} \ , \\
    T_2 &= \frac{0.51 \beta_0^2}{\left(1 + 0.90 d^2 + 0.69 \beta_0^{12/5}\right)^{5/6}} \ ,
\end{align}
where
\begin{equation}
    \beta_0 = 0.4065\sigma_R^2
\end{equation} represents the spherical wave Rytov variance and 
\begin{equation}
    d = \sqrt{\frac{kD_{\rm Rx}^2}{4L}} \approx 3 \ . 
\end{equation}

We report in Fig.~\ref{fig:simulations}a the simulation of the coupling efficiency due to scintillation expected in our experiment, as a function of the Fried parameter $r_0$. 
In the simulation, the Fried parameter ranges between $3$ and $15$~cm at the design wavelength of $1555$~nm. 
These values have been chosen on the basis of previous measurements conducted over the channel, in a way similar to the one described in Section~\ref{sec:AO_results} for analyzing the WFS data.
From Fig.~\ref{fig:simulations}a it is clear that, even in the worst case of a Fried parameter $r_0$ of the order of a few centimeters, the detrimental contribution of scintillation to the SMF coupling efficiency reamins contained within $-1$~dB. 

The final factor influencing coupling efficiency, as described in Eq.~\eqref{eq:eta_SMF}, is determined by the ability of the AO system to correct phase perturbations on the wavefront. 
As shown in Eq.~\eqref{eq:eta_AO}, this contribution ($\eta_{\rm AO}$) can be further decomposed into two distinct components that quantify the spatial ($\eta_{\varphi}$) and the temporal ($\eta_{\tau}$) correction efficiencies of the AO system. 
Moreover, spatial efficiency itself can be expanded as detailed in Eq.~\eqref{eq:eta_varphi}, taking into account the effective terms corrected by the AO system ($\eta_{\varphi, {\rm ON}}$) and the wavefront residual ($\eta_{\varphi(J)}$). 
The standard AO approach is based on the decomposition of the instantaneous phase perturbation $\varphi$ at the entrance pupil of the telescope in terms of Zernike standard coefficients $Z_j(\rho, \theta)$ as
\begin{equation}
    \varphi(\rho, \theta) = \sum_{j=1}^{\infty} b_j Z_j(\rho, \theta)\ ,
\end{equation}
where $(\rho, \theta)$ are the normalized pupil coordinates, and the Zernike modes are indexed by $j$, following the ANSI/ISO standard convention to represent different aberration orders. 
As demonstrated by Noll in his work~\cite{noll1976zernike}, Zernike polynomials can be used to represent the Kolmogorov specturm of atmospheric turbulence, with the associated coefficients modeled as zero-mean Gaussian random variables. 
From this result, Boremann~\cite{boremanZernikeExpansionsNonKolmogorov1996} derived an expression for the variance of Zernike coefficients due to turbulence given by Eq.~\eqref{eq:eq_fit}, with 
\begin{equation}
    g(j) =  \frac{n+1}{\pi}  \frac{ \Gamma \left( n-\frac{5}{6} \right) \Gamma \left(\frac{23}{6}\right) \Gamma \left(\frac{11}{6}\right) \sin \left(  \frac{5}{6}\pi \right)}{ \Gamma \left(n+\frac{23}{6}\right)} \ ,
\end{equation}
where
\begin{equation}
    n = \left\lceil \frac{-3 + \sqrt{9+8j}}{2} \right\rceil 
\end{equation}
denotes the radial order of the specific aberration \cite{Schwiegerling2014}.
When modelling AO systems, we consider a system that is able to reduce the variance term of one or more aberration orders. 
In particular, for an ideal AO system that perfectly compensates mode variances up to the $J$-th Zernike order, the residual phase variance $\sigma_J^2$ can be estimated via~\cite{noll1976zernike}
\begin{equation}
    \sigma_J^2 = 0.2944  J^{-\sqrt{3}/2}  \left( \frac{D_{\rm Rx}}{r_0}\right)^{5/3} \ , \label{eq:sigmaj2}
\end{equation}
from which the coupling efficiency $\eta_\varphi \equiv \eta_{\varphi(J)} $ due to spatial phase variance can be written as
\begin{equation}
    \eta_{\varphi(J)} = e^{-\sigma_J^2} \ . \label{eta:spatial:J}
\end{equation}
We report in Fig.~\ref{fig:simulations}b the simulation of the efficiency term $\eta_{\varphi(J)}$ obtained by assuming ideal correction of different $J$ modes. 
In particular, according to the actuators of the DMF at our disposal, we expect to be able to correct (at least partially) the first 35 aberration modes~\cite{tyson2022principles}. 
Therefore, $J=35$ represents the baseline for our implementation, and it is easy to check that for a Fried parameter ranging from $3$ to $15$~cm, the expected spatial efficiency due to the residual phase variance ranges from $-5$ to $-0.5$~dB, while it cannot be better than $-4$ dB with only tip-tilt correction ($J=2$). 

The temporal characteristics of atmospheric turbulence and the AO system response affect the coupling efficiency through the factor $\eta_\tau$, defined as
\begin{equation}
    \eta_\tau = e^{-\sigma_\tau^2} \ , \label{eq:etatau}
\end{equation}
where
\begin{equation}
    \sigma_\tau^2 = \left( \frac{f_G}{f_{\rm 3dB}}\right)^{5/3}  \label{eq:sigmatau2}
\end{equation}
is the temporal phase variance given in terms of the Greenwood frequency $f_G$ and of the rejection bandwidth of the AO system $f_{\rm 3dB}$. 
For a horizontal link, the Greenwood frequency can be related to the Fried parameter and the mean wind speed (along the channel) $w$ by~\cite{tyson2022principles}
    \begin{equation}
        f_G = 0.43 \frac{w}{r_0} \ . \label{eq:greenwood}
    \end{equation}

In our system, the $3$~dB control bandwidth is limited to $10$~Hz. 
We report in Fig.~\ref{fig:simulations}c the simulation of the efficiency term $\eta_\tau$ obtained by supposing different wind speeds and a $3$~dB control bandwidth of $f_{\rm 3dB} = 10$~Hz. 
It is evident that for our implementation the AO control bandwidth is the major limitation to reach a better SMF-coupling efficiency. 
However, such limitation can be overcome by exploiting dedicated real-time controllers (RTC), which can reach control bandwidths of more than $100$~Hz. 

\subsection{Detailed description of the transmitter node}
\label{sec:TX_node}

The Tx node is located at the summit of Monte Grande ($45^\circ21'42.4''$~N $11^\circ40'21.7''$~E, $474$~m a.s.l.) in the Colli Euganei regional park, and is installed on a weather station of the Agenzia Regionale per la Prevenzione e Protezione Ambientale del Veneto (ARPAV). The site on top of Monte Grande is shown in Fig.~\ref{fig:setup}b and consists of a management room and a radar tower for weather monitoring. 
A military-graded SMF is used to connect the management room, where the QKD source is located, to a custom-built mechanical enclosure, which is installed on the balcony of the meteorological tower and hosts the optical transmitter.

Inside the enclosure, a filtering stage exploiting wavelength-division multiplexers (WDMs) allows to combine the QKD signal with additional beacon lasers. 
In particular, the quantum signal is combined with another C-band beacon laser at $\lambda_{\rm Beacon} = 1545.32$~nm (channel 40 of ITU-grid), that is used to provide feedback for the AO system implemented at the receiver side, and with another reference signal at 850~nm which is used for coarse alignment. 
After the WDMs, a 90:10 fiber beam splitter is added before a compact transmitter module, similar to the one described in Ref.~\cite{piccia_DFA_DEI}.
The lower output port is connected to a fiber-coupled PM to monitor the output power, while the higher output power port is connected to the transmitter module, that generates a collimated output beam of waist $W_0=25$~mm for the two  C-band signals. 
The optical signals are transmitted from the enclosure through an optical window with a clear aperture of 150~mm pointing towards the Rx node located in the city of Padova.
The QKD source is a customized version of the commercial product QUKY-TX, that allows to tune the amplitude level of the emitted pulses by exploiting a variable optical attenuator (VOA), in order to set the correct mean photon number at the optical window.
To enable coarse alignment with the Rx terminal, the optical transmitter is mounted on an alt-azimuth mount equipped with a system of lenses 
that images the scene on a coarse camera (CAM), providing a full-angle field-of-view (FOV) of $8$~mrad. 

\subsection{Detailed description of the receiver node}
\label{sec:Rx_node}

The Rx node --- see Fig.~\ref{fig:setup}c --- is located in the city of Padova and is comprised of two buildings of the University; DEI ($45^\circ24'33.1''$~N, $11^\circ53'39.9''$~E, $12$~m a.s.l.) and Luxor ($45^\circ24'29.0''$~N, $11^\circ53'48.5''$~E), which are linked by a deployed fiber of about 0.5~km. 
The OGS consists of a dome and an equatorial mount that supports the telescope.

The telescope used in the testbed is an $f/8$ Ritchey-Chrétien mounted on a German Equatorial Mount (model Paramount ME-II by Software Bisque), and is equipped on the top flange with an additional coarse telescope (model SkyMax 127 by Sky-Watcher) for coarse pointing towards the Tx node. 
The coarse telescope is an $f/12$ Maksutov with an aperture of 127~mm, and is equipped with an acquisition camera (model ASI 294 MC Pro Color by ZWO) with a full-angle FOV of $9 \times 13$~mrad$^2$. 
On its rear, the receiving telescope is equipped with a mechanical breadboard of $600\times600$~mm$^2$, where the AO system is installed.

The AO system is based on the usage of a DFM, driven in closed-loop by the feedback of a Shack-Hartmann WFS. For this reason, the incoming aberrated C-band signals are split using a 90:10 beam splitter, where $10\%$ of the optical power is reflected toward the WFS for wavefront measurement, and the remaining $90\%$ is transmitted through the beam splitter.
The transmitted C-band signals then propagate through a bulk band-pass filter centered at $1550$~nm with a bandwidth of $40$~nm full width at half maximum (FWHM) before being coupled into the SMF via a coupling lens.
After fiber injection, a dense wavelength-division multiplexer (DWDM) with $100$~GHz spacing centered at $\lambda_{\rm QKD}$ is used to separate the C-band beacon from the QKD signal. 
In particular, the C40 beacon is going to be measured with a fiber-coupled PM that is used to monitor the channel efficiency, while the C15 quantum signal is directed to the deployed fiber connecting DEI to Luxor. 
At Luxor, we use a commercial QUKY-RX EduPro as QKD receiver, that allows to choose either the internal InGaAs-based SPADs or different external detectors, such as SNSPDs.

The DFM used in our setup is a piezoelectric bimorph mirror with a clear aperture of 22~mm and 64 actuators arranged in a radial pattern (model DM6422). 
The  WFS (model SWIR640) is based on an InGaAs sensor of size $9.6 \times 7.7$~mm$^2$ which features a lenslet array of 250~$\mu$m pitch size with 15~$\mu$m pixel size, allowing a maximum full-frame rate of 600~Hz. 

To correct the turbulence-induced aberrations of the received wavefront, the DFM and WFS need to be conjugated to the aperture pupil of the telescope. 
Another constraint is that the size of the beam has to be small enough to be manipulated with 1" optics, and also magnified appropriately to cover as much of the DFM aperture as possible, so all actuators are exploited. 
With these requirements in mind, 
a series of optical relays allows to obtain three replica of the telescope entrance pupil (yellow stars in Fig.~\ref{fig:setup}c) along the breadboard. The second pupil, with a diameter dimension of 21~mm is used to locate the DFM, while the third one (with a diameter of 6.25~mm) is used to locate the WFS.

To ensure proper operation of the DFM, it is necessary to perform a calibration of the system through the characterization of the DFM's interaction matrix~\cite{chiuso_calibration}. The interaction matrix permits to estabilish the relation between each DFM's actuators and the wavefront slopes measured by the WFS.
This procedure is essential for enabling the control software to selectively activate the appropriate actuators to compensate for the incoming aberrated wavefront, and also to correct for static aberrations that are intrinsic to the optical system. For this purpose, a flip mirror (FLM) is placed between the collimating lens and the first pupil plane, allowing injection of a reference collimated beam into the AO bench. This beam, generated by a calibration laser (CL) at the same wavelength as $\lambda_{\rm Beacon}$, is matched in diameter to the incoming beam, and is used to calibrate the DFM via PhotonLoop.

\section*{ACKNOWLEDGEMENTS}
\label{sec:ack}

We would like to acknowledge for the support: Agenzia Spaziale Italiana (2020-19-HH.0 CUP Grant No. F92F20000000005, Italian Quantum CyberSecurity I-QKD), Horizon Europe Project “Quantum Secure Networks Partnership” (QSNP, Grant agreement No. 101114043). 
I. Karakosta-Amarantidou acknowledges funding from the European Union’s Horizon 2020 research and innovation programme under the Marie Skłodowska-Curie Grant Agreement No. 956071 AppQInfo-Applications And Hardware For Photonic Quantum Information Processing. 
The authors would like to aknowledge L. Dal Lago, M. Previato, G. Cenzon, M. Dianin of ARPAV for the collaboration and support at Monte Grande, as well as F. Bettini, F. Luise and the Security Personnel of DEI for the collaboration and support at the department.

{\it Disclaimer.} Funded by the European Union. Views and opinions expressed are however those of the author(s) only and do not necessarily reflect those of the European Union or European Commission-EU. Neither the European Union nor the granting authority can be held responsible for them.

\section*{Author contributions}
E.R., I.K.A., M.P., M.N. and F.V. contributed in the realization of the free-space optical link.
A.V. and S.B. supported the implementation of the adaptive optics part of the setup.
E.R., M.P., M.N., M.A., F.B.L.S., M.T. and F.V. performed the QKD experiments.
G.V., P.V. and F.V. supervised the entire work.
All authors participated in the discussion of results and contributed to the final manuscript.

\end{document}